\documentclass[reprint, 
    aps,prl,
    superscriptaddress,
    longbibliography,
    amsmath,amssymb,
    balancelastpage
]{revtex4-2}

\usepackage[utf8]{inputenc}
\usepackage[T1]{fontenc}
\usepackage{graphicx}   
\usepackage{dcolumn}    
\usepackage{bm}         

\usepackage{mathptmx}
\usepackage{etoolbox}
\usepackage{siunitx}
\usepackage{ORCIDinREVTeX}
\usepackage{xcolor}
\usepackage{marvosym}

\usepackage[unicode=true,
    colorlinks=true,
    linkcolor=blue,         
    citecolor=blue,         
    filecolor=blue,         
    urlcolor=blue]{hyperref}

\renewcommand*{\section}[1]{\paragraph{#1}}
\renewcommand*{\eqref}[1]{\hyperref[{#1}]{\textup{(\ref*{#1})}}}
\newcommand*{\figref}[1]{\hyperref[{#1}]{\textup{Fig.~\ref*{#1}}}}
\newcommand*{\secref}[1]{\hyperref[{#1}]{\textup{Sec.~\ref*{#1}}}}
\newcommand*{\tabref}[1]{\hyperref[{#1}]{\textup{Table~\ref*{#1}}}}
\newcommand*{\vect}[1]{\mathbf{#1}}  

\begin{document}
\title{Clock Pulling Enables Maximum-Efficiency Wireless Power Transfer}

\author{Xianglin~Hao\textsuperscript{\Letter}}  
\thanks{These authors contributed equally to this work.\\
\hspace*{-1em}\Letter Author to whom correspondence should be addressed. Please e-mail to:  \href{mailto:xianglhao2-c@my.cityu.edu.hk}{xianglhao2-c@my.cityu.edu.hk}}
\orcid{0000-0001-7149-0113}

\author{Xiaosheng~Wang}
\thanks{These authors contributed equally to this work.\\
\hspace*{-1em}\Letter Author to whom correspondence should be addressed. Please e-mail to:  \href{mailto:xianglhao2-c@my.cityu.edu.hk}{xianglhao2-c@my.cityu.edu.hk}}
\orcid{0000-0002-5167-3972}
\affiliation{Department of Electrical Engineering, City University of Hong Kong, Hong Kong, 999077, China}

\author{Ke~Yin}
\orcid{0000-0002-8534-216X}
\affiliation{College of Electronics and Information Engineering, Sichuan University, Chengdu, 610101, China}%

\author{Sheng~Ren}
\orcid{0000-0003-3181-9265}

\author{Chaoqiang~Jiang}
\orcid{0000-0001-5374-364X}
\affiliation{Department of Electrical Engineering, City University of Hong Kong, Hong Kong, 999077, China}

\author{Jianlong~Zou}
\orcid{0000-0003-1489-0828}

\author{Tianyu~Dong}
\orcid{0000-0003-4816-0073}
\affiliation{School of Electrical Engineering, Xi’an Jiaotong University, Xi’an 710049, China}%

\author{Chi~Kong~Tse}
\orcid{0000-0002-0462-3999}
\email{chitse@cityu.edu.hk}
\affiliation{Department of Electrical Engineering, City University of Hong Kong, Hong Kong, 999077, China}

\date{\today}

\begin{abstract}
Nonlinear parity-time (PT) symmetry in non-Hermitian wireless power transfer (WPT) systems, while attracting significant attention from both physics and engineering communities, have posed formidable theoretical and practical challenges due to their complex dynamical mechanisms. Here, we revisit multistability in nonlinear non-Hermitian systems and find that the PT-symmetry state is not always stable even in PT-symmetry phase.
We report a discovery on a nonlinear clock-pulling mechanism, which can forcibly break the PT symmetry. Proper implementation of this mechanism can switch the system stability, particularly in stabilizing the conventional unstable state which has the maximum transfer efficiency for WPT. Our work offers new tools for non-Hermitian physics and is expected to drive technological progress.
\end{abstract}

\maketitle 

\section{Introduction}
Near-field wireless power transfer (WPT) technology has experienced rapid development in the past two decades and has already seen small-scale commercial applications with continued growth \cite{kurs2007wireless,hui2013critical,assawaworrarit2017robust,krasnok2018coherently,jiang2019fractional,vu2022operation}. However, the field lacks a consensus on foundational design principles for wireless power transfer, as is often the case with nascent electronic technologies. Typically, researchers can only struggle to explore the improvement space to optimize performance based on existing excellent converter topologies and various control methods, leading to the dilemma of balancing system performance and design complexity. This predicament arises primarily from a long-standing lack of intuitive and effective understanding of the underlying physics of the WPT system, specifically the physical mechanisms governing coupled-resonator systems. Although numerous theories suggest that coupled-resonator systems can achieve maximum power transmission and optimal efficiency at specific frequencies \cite{hui2013critical,hua2022autonomous}, the mechanisms to fabricate and maintain systems at this optimal frequency under perturbations remain unclear. Due to insufficient understanding of the underlying physical mechanisms, designers resort to exploring complex optimization methods to sustain high-efficiency power transfer; however, such approaches often prove over-ideal and impractical.

Recently, the non-Hermitian scheme \cite{zeng2022efficient,song2021wireless,ashida2020non,moccia2020line,xiao2019enhanced,el2018non,wu2022generalized,hao2023frequency}, especially the emergence of the parity-time symmetric WPT system \cite{assawaworrarit2017robust,assawaworrarit2020robust,zhou2019nonlinear}, has provided a direction for robust wireless power transfer at self-oscillation frequency \cite{assawaworrarit2017robust}, which provides a means of self-maintaining frequency under disturbances. However, the frequency of the PT symmetrical phase is not optimal with maximum efficiency and transmission power. Furthermore, the multimode characteristics and fundamental constraints inherent to nonlinear non-Hermitian WPT systems remain poorly understood, with their practical exploitation still constituting a formidable open challenge.

In this paper, we investigate the dispersion relationship of the nonlinear gain that has long been overlooked in PT-symmetric systems, and demonstrate for the first time that forced symmetry breaking of the PT symmetric phase in nonlinear PT-symmetry systems is also possible. Our previous work demonstrated the use of dispersive gain designs to select the steady states of non-Hermitian WPT systems at asymmetric resonance \cite{hao2025dispersive}. Here, we further clarify that while PT-symmetric systems maintain dispersion-independent steady-state gain, their neighboring dispersion characteristics in the frequency domain can be adjusted to achieve steady-state selection. We reveal how a nonlinear clock pulls the non-Hermitian system to a specific steady state, steering dynamics toward symmetry-broken states, which we term forced symmetry breaking. This fundamental insight enables our experimental realization of non-radiative WPT with a theoretical maximum efficiency.

\vspace{1em}

\section{Non-radiative WPT systems}
\begin{figure}[!t]
    \centering
    \includegraphics{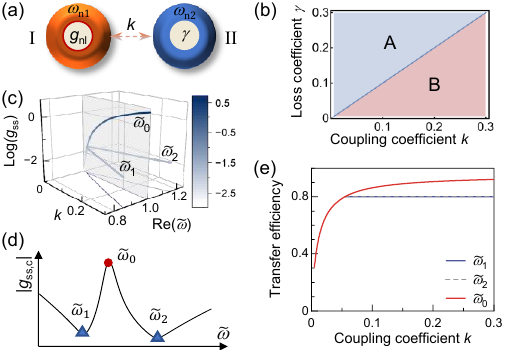}
    \caption{(a) Schematics of a coupled-resonator dimer with a nonlinear gain $g$ and loss $\gamma$. (b) Phase diagram of the asymmetric resonance dimer. (c) Comparison of steady-state gain $g_\text{ss}$ of the different states. (d) Steady-state requried gain value $g_\text{ss}$ and complex eigenfrequencis evolution versus coupling parameter $k$ of the parity-time symmetry system. (e) Theoretical transfer efficiency versus the coupling parameter $k$. Here, $\omega_\text{n1}$ and $\omega_\text{n2}$ denote the natural resonance frequency of the resonators I and II; $k$ denotes the coupling coefficients between resonators; $\widetilde{\omega}_1$, $\widetilde{\omega}_2$, and $\widetilde{\omega}_0$ represent three theoretical steady-state modes of the system. For all figures, $\gamma=0.0565$, while the intrinsic loss of the two resonators are $\gamma_\text{s}=1/95$ and $\gamma_\text{r}=1/380$.}
    \label{fig:fig01}   
\end{figure}

We start by analyzing the steady-state mechanism of the two-coil WPT system \cite{assawaworrarit2017robust,zhou2019nonlinear}, as shown in \figref{fig:fig01}(a). To broaden the theoretical applicability, we consider the scenario where the resonators are constructed with an arbitrary ratio of resonant frequency parameters, \textit{i.e.}, $\omega_\text{n1} =\chi_c^2 \chi_l^2 \omega_\text{n2}$, where $\chi_c$ and $\chi_l$ denote the proportionality coefficients of capacitance and inductance, respectively \cite{supplementary}. Using the coupled-mode theory (CMT) \cite{haus1991coupled,assawaworrarit2017robust}, the time evolution of the amplitudes of the transmitter and receiver resonator for the generalized dimer \cite{schindler2011experimental,schindler2012symmetric,assawaworrarit2017robust,assawaworrarit2020robust}, denoted by $\vect{a} = [a_1,a_2]^\text{T}$, is governed by $ -i \frac{\text{d}\vect{a}}{\text{d}t} =  H_\text{gad} \vect{a}$
where
\begin{equation}\label{eq:heff}
    H_{\text{gad}}=\omega_\text{n2} \left(
\begin{array}{cc}
     \chi_c^2 \chi_l^2 - i  \frac{g_\text{nl}(a_1) -\gamma_\text{s}  }{2 \chi_c^2 } & -\frac{k }{2 \chi_c \chi_l} \\
 -\frac{k }{2 \chi_c \chi_l}  & 1 + i  \frac{\gamma }{2}
\end{array}
\right).
\end{equation}
Here, $g_\text{nl}(a_1)$ describes the strength of the gain in the transmitter resonator, which is a nonlinear function of the normal mode $a_1$ and depends on the design of the gain element, while $\gamma_\text{s}$ is the inherent loss of the source resonator. The total dissipation parameter $\gamma$ of the receiving resonator comprises both load loss $\gamma_\text{l}$ and intrinsic loss $\gamma_\text{r}$, expressed as $\gamma=\gamma_\text{l}+\gamma_\text{r}$. Also, $k$ denotes the coupling parameter between the resonators, which comes from the mutual inductance between the coils in this paper. Such a Hamiltonian can form a PT-symmetric system when $\chi_c= \chi_l = 1$ \cite{assawaworrarit2020robust}. To analyze the steady-state characteristics, we linearize \eqref{eq:heff} by assuming that the system will reach a steady state under a specific gain value $g_\text{ss}$, \textit{i.e.}, $g_\text{nl}(a_1)-\gamma_\text{s} \to g_\text{ss}$ in the steady state. Based on this assumption, to determine the steady-state frequency and the requiring gain, one can get the characteristic equation by solving $\text{Det}\left(\widetilde{\omega} \vect{I} - H_\text{gad}/\omega_\text{n2} \right) = 0$ (where $\vect{I}$ denotes an identity matrix and $\widetilde{\omega}$ denotes the normalized frequency), yielding
\begin{align}
    \widetilde{\omega} ^2 -(1+\chi_c^2 \chi_l^2) \widetilde{\omega} +\chi_c^2 \chi_l^2 +\frac{1}{4 \chi_c^2} (\gamma g_\text{ss} -\frac{k^2}{\chi_l^2}) + \notag \\
    i \left[\frac{1}{2} \widetilde{\omega}  (\frac{ g_\text{ss}}{\chi_c^2} -\gamma)+\frac{1}{2}  (\chi_c^2 \chi_l^2 \gamma - \frac{ g_\text{ss}}{\chi_c^2})\right] =0 \label{eq:chara}
\end{align}
Let the real and imaginary parts of \eqref{eq:chara} be 0, we have
\begin{subequations}
    \begin{align}
    g_\text{ss} &= \gamma \chi_c^2 \frac{\widetilde{\omega} -\chi_c^2 \chi_l^2}{\widetilde{\omega} -1}, \label{eq:3a} \\
    \widetilde{\omega} ^2 - (1+\chi_c^2 \chi_l^2) \widetilde{\omega} +\chi_c^2 \chi_l^2 &= \frac{k^2}{4 \chi_l^2 \chi_c^2}  -\frac{\gamma  g_\text{ss}}{4 \chi_c^2}. \label{eq:3b}
\end{align}
\end{subequations}

According to \eqref{eq:3a}, we can determine the steady-state requiring gain $g_\text{ss}$ for different modes. By combining \eqref{eq:3a} and \eqref{eq:3b}, one can compute eigenfrequencies \cite{supplementary}. It is worth mentioning that the steady-state gain $g_\text{ss}$ is a function of the frequency from \eqref{eq:3a} \cite{hao2025dispersive}. However, when the system satisfies PT symmetry with $\chi_l =\chi_c=1$, the frequency-dependent feature will disappear and \eqref{eq:3a} degenerates into $g_\text{ss}=\gamma$. This frequency-independent gain supports two theoretical states $\widetilde{\omega}_{1,2}=1\mp \frac{1}{2} \sqrt{k^2-\gamma^2}$. Besides, in the PT-symmetry system, the imaginary part of \eqref{eq:chara} will also be zero at $\widetilde{\omega}_0=1$. To support the mode $\widetilde{\omega}_0=1$, the steady-state gain must satisfy $g_\text{ss}=k^2/\gamma$. For any $k$, the state $\widetilde{\omega}_0$ always exists, but only when $k>\gamma$, $\widetilde{\omega}_{1,2}$ are real states. Therefore, there are two different parameter systems in the system, as shown in \figref{fig:fig01}(b). The system has three different states in Region A (PT-symmetry phase) and only one state $\widetilde{\omega}_0$ in Region B (PT-broken phase). 

In the PT-symmetry phase of known non-Hermitian systems, only $\widetilde{\omega}_{1,2}$ has been observable so far, with state $\widetilde{\omega}_0$ is claimed to be unstable \cite{assawaworrarit2017robust,zhou2019nonlinear,zeng2022efficient,song2021wireless}. This phenomenon typically is explained by the principle of minimal gain, i.e., only the state with the lowest gain remains stable due to gain saturation. As shown in \figref{fig:fig01}(c), state $\widetilde{\omega}_0$ requires the highest gain as $k$ varies in PT-symmetry phase, and thus has been regarded as an unstable state. In fact, the state $\widetilde{\omega}_0$ is also an extremum point of the gain, which can be demonstrated by adopting a different approach to get the steady-state gain. If the gain $g_\text{ss}$ is not assumed to be a real number, we can directly find the steady-state required complex gain $g_\text{ss,c}$ from \eqref{eq:chara}, which yields
\begin{align}\label{eq:gssc}
    g_\text{ss,c}&= \frac{ k^2 \gamma }{ \chi_l^2 \left(\gamma ^2+4 (\widetilde{\omega} -1)^2\right)} + \notag \\
    &i  \left(2 \chi _c^2 \widetilde{\omega} +\frac{2k^2 (\widetilde{\omega} -1)}{ \chi_l^2 \left(\gamma ^2+4 (\widetilde{\omega} -1)^2\right)} -2 \chi _c^4 \chi _l^2 \right)
\end{align}
It is evident that gain $|g_\text{ss,c}|$ has a relative extremum at $\widetilde{\omega}=\widetilde{\omega}_0=1$, as shown in \figref{fig:fig01}(d). However, an extremum point does not necessarily imply instability. As an illustrative example, consider the motion of a ball on the potential energy surface formed by the curve in \figref{fig:fig01}(d). By implementing a real-time responsive pulling mechanism—one that pulls the ball leftward when it tends to roll rightward past the peak, and rightward when it tends to roll leftward—the ball can be stabilized at the peak point. For the non-Hermitian circuit system, this pulling effect can be achieved using feedback control, provided the feedback response is faster than the frequency variation of the system. However, experimental studies have shown that the transient frequency shifts in non-Hermitian systems can be extremely rapid and discontinuous \cite{schindler2011experimental,assawaworrarit2017robust,zhang2018parity,hao2025dispersive}. This quantum-like dynamical behavior introduces significant challenges in frequency stabilization. To overcome this, we introduce a classical clock in the feedback loop to enforce continuous frequency variation. Subsequently, feedback control can be applied to steer the non-Hermitian system's frequency, enabling stabilization at $\widetilde{\omega}_0$. Crucially, stabilizing the system at $\widetilde{\omega}_0$ is not only physically intriguing but also of practical importance. As illustrated in \figref{fig:fig01}(e), the state  consistently exhibits the highest power transfer efficiency. If a non-Hermitian system can be locked to $\widetilde{\omega}_0$, it would substantially enhance the transfer performance of the robust WPT.

\begin{figure}[!ht]
    \centering
    \includegraphics{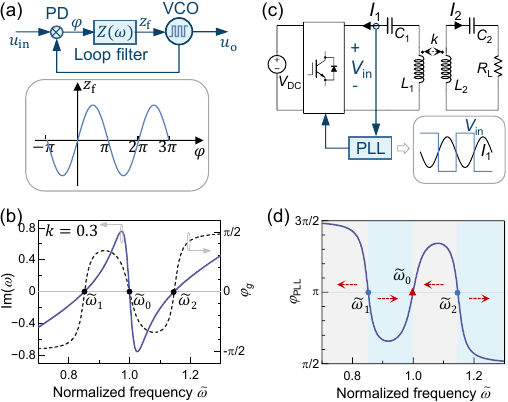}
    \caption{(a) Schematic diagram of the phase-locked loop (PLL) and its phase detection characteristics. (b) The dispersion curves of the imaginary part Im$(g_\text{ss,c})$ and phase angle $\varphi_g$ of the steady-state required gain $g_\text{ss,c}$ when $\chi_l=\chi_c=1$. (c) A simplified schematic of a WPT system based on the clock-pulling gain module using the PLL shown in (a). (d) Dispersion relation of the steady state required gain for the PT-symmetric system based on the gain module in (c). Red arrows indicate the direction of dynamic evolution.}
    \label{fig:fig02}
\end{figure}

In electronic systems, phase-locked loops (PLLs) can be used to achieve clock pulling. As shown in \figref{fig:fig02}(a), the voltage-controlled oscillator (VCO) within the PLL provides a classical clock that enforces continuous frequency variations through a feedback control \cite{gardner2005phaselock}. Also, \figref{fig:fig02}(a) illustrates the phase-detection characteristics of the PLL based on sinusoidal phase detectors (PD), where $\varphi$ is the phase difference between input signal $u_\text{ref}$ and output signal $u_\text{o}$, with $z_\text{f}$ denoting the feedback quantity for regulating the frequency variation of VCO. When $z_\text{f} > 0$, the frequency of  $u_\text{o}$ increases, whereas for $z_\text{f} < 0$, it decreases. The PLL exhibits two types of stable operating points. In the first type ($\varphi=(2n+1)\pi$ where $n$ is an integer), the output signal $u_\text{o}$ is in phase with $u_\text{ref}$, and $z_\text{f}$ varies positively with phase deviations nearby. In contrast, in the second type ($\varphi=2n\pi$ where $n$ is an integer), the $u_\text{o}$ is anti-phase with the $u_\text{ref}$ while the phase-detector response becomes negative around this stable point. This dual stability feature enables different clock-pulling polarities for frequency control. 

With the continuous frequency variation, we can evaluate the influence of clock pulling on the stability of the gain module through the dispersion relation from \eqref{eq:gssc}. Here, \figref{fig:fig02}(b) shows the frequency-dependence of the imaginary part of the gain Im($g_\text{ss}$), where the gain is also complex-valued except at three steady-state modes. Moreover, state $\widetilde{\omega}_0$ displays reversed Im($g_\text{ss}$) polarity in its adjacent frequency domains relative to the other two states. This produces a similar phase response as shown in \figref{fig:fig02}(b), where the left neighborhood exhibits negative phase angles versus positive angles on the right. Proper design of the pulling polarity based on these characteristics may enable stabilization of state $\widetilde{\omega}_0$ as a robust monostable state, unaffected by system symmetry constraints. Also, \figref{fig:fig02}(c) presents the PLL-based clock-pulling system designed to stabilize $\widetilde{\omega}_0$. A current sensor samples the current $I_1$ as the PLL reference, with the switch network outputting a voltage $V_\text{in}$ in phase with the PLL output signal $u_\text{o}$. Thus, the phase detector output corresponds to the current-voltage phase difference in ideal case from \figref{fig:fig02}(a), i.e., $\varphi_\text{PLL}=\text{Arg}(I_1/V_\text{in})= \text{Arg}(-1/g)$. 

The phase-frequency relationships of the PLL is shown in \figref{fig:fig02}(d). Although the PLL exhibits two types of fixed points at 0 and $\pi$, the dynamics of non-Hermitian systems necessitates negative resistance to provide gain, thereby restricting the equilibrium point of the system to $\varphi_\text{PLL}=\pi$, as indicated by the gray horizontal line in \figref{fig:fig02}(d). Away from equilibrium, the PLL adaptively adjusts the output frequency based on the phase difference $\varphi_\text{PLL}$, thereby pulling the frequency evolution of the non-Hermitian system. The red arrows in \figref{fig:fig02}(d) illustrates how the clock pulling stabilize the state $\widetilde{\omega}_0$: Benefiting from the negative frequency-phase polarity when the PLL operates near the point $\varphi_\text{PLL}=\pi$, the system frequency increases continuously in the light-blue band ($\varphi_\text{PLL}<\pi$) and decreases continuously in the light-green band ($\varphi_\text{PLL}>\pi$). This frequency-phase feedback selectively stabilizes the state $\widetilde{\omega}_0$ while destabilizing $\widetilde{\omega}_1$ and $\widetilde{\omega}_2$. Thus, the clock pulling reconfigures the stability, rendering $\widetilde{\omega}_0$ uniquely robust stable even in the PT-symmetry phase. In particular, the scheme exploits dynamical frequency-phase coupling rather than modifying the dispersion landscape itself. Furthermore, even when PT symmetry is not satisfied, the phase-frequency response of the non-Hermitian system's gain remains similar to \figref{fig:fig02}(d), except for slight variations in the zero-point value. Our clock-pulling scheme can stabilize the zero-point $\widetilde{\omega}_0$ in asymmetric systems where $\chi_l\chi_c \neq1$ as well.


\begin{figure*}[!t]
    \centering
    \includegraphics[width=5.5in]{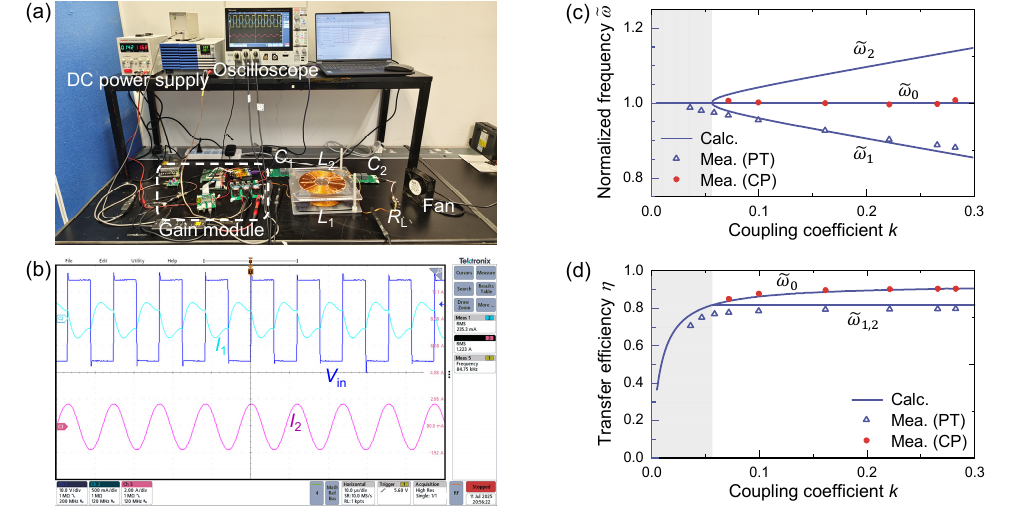}
    \caption{(a) Photo of the experimental setup. (b) Measured steady-state waveforms of the WPT system based on clock-pulling gain when $k=0.28$. (c) Normalized steady-state frequency and (d) power transfer efficiency , both plotted versus the coupling coefficient $k$. Here, the red circles denote measurement results for the proposed clock-pulling (CP) scheme, while the blue hollow triangles represent conventional parity-time-(PT)-symmetric system measurements. The loss parameter $\gamma=0.0565$ for all the figures.}
    \label{fig:fig03}
\end{figure*}

To verify our method, we prototyped a test system based on the clock-pulling gain, as shown in \figref{fig:fig02}(c). The switch network, using full-bridge inverters in this paper, generates a square-wave voltage $V_\text{in}$ that is anti-phase with current $I_1$ under PLL regulation, thereby providing a nonlinear gain. The PLL is realized using the high-performance field programmable gate array chip EP4CE10F17, as detailed in \textbf{S3} of the Supplementary Material \cite{supplementary}. Two planar coils (24 turns of Litz wire, $\approx \SI{71}{\micro\henry}$ each) were used as resonators, with high-quality capacitors tuning the resonant frequency to 85.13~kHz. Experimental tests were conducted with a load resistor $R_\text{L}=\SI{2}{\ohm}$ and voltage $V_\text{DC}=\SI{15}{\volt}$, while the coupling coefficient was varied by adjusting the offset of the coils. 

\figref{fig:fig03}(a) shows the photo of the experimental setup, with detailed parameters provided in the Supplementary Material. \figref{fig:fig03}(b) show the measured waveforms of the input voltage $V_\text{in}$ and the resonator currents $I_1$ and $I_2$ for $k = 0.28$. It is evident that the system operates stably at 84.75~kHz, which shows excellent agreement with the theoretically predicted frequency of 85.13~kHz, confirming that the proposed method successfully stabilizes mode $\widetilde{\omega}_0$. Although the nonlinearity of the negative resistance introduces non-negligible high-order harmonics and brings some distortion to the primary-side current waveform, it basically does not affect the steady-state performance at the fundamental frequency. The steady-state waveform with distortion can be calculated by considering the harmonics \cite{supplementary}. 

\figref{fig:fig03}(c) and (d) show the normalized frequency and efficiency versus the coupling coefficient $k$, demonstrating excellent agreement between the measurement and the calculation based on the CMT model. In the so-called PT symmetric phase (the blank region in \figref{fig:fig03}(c) and (d)), by applying clock pulling, the non-Hermitian system can spontaneously stabilize to the asymmetric state $\widetilde{\omega}_0$. This phenomenon unveils richer dynamics in non-Hermitian systems, suggesting the possible existence of undiscovered interaction mechanism of nonlinearity and symmetry. The shaded region corresponds to the PT-broken phase, where only a single real eigenfrequency $\widetilde{\omega}_0$ exists and the gain phase exhibits strict positive correlation with frequency. Clock-pulling systems configured per \figref{fig:fig03} is unstable in this regime, as detailed in \textbf{S2} of the Supplemental Material \cite{supplementary}. Experimentally observed PT-symmetric states exhibit slight deviations from theoretical predictions due to minor asymmetries in the system. The measured results also demonstrate remarkable practical potential. As shown in \figref{fig:fig03}(c) and (d), the system exhibits stable frequency and high transfer efficiency. The clock-pulling scheme achieves the theoretically maximum transfer efficiency for two-coil systems without active tuning over varying, consistently outperforming parity-time-symmetric approaches in the strong coupling region ($k>\gamma$).

\section{Conclusion}
In this paper, we analyze the wireless power transfer characteristics in non-Hermitian systems and report the findings of a steady-state mechanism driven by clock pulling, which can forcibly break the parity-time symmetry. Our work has established an intuitive physical picture of the steady state selection mechanisms enabled by  clock pulling. In multi-mode non-Hermitian systems, clock pulling can switch the steady-state frequency, enabling the stabilization of conventionally unstable states via feedback design. Taking the PT-symmetric system as an example, this work demonstrates the frequency-selective effect of clock pulling, which can be equally applied to asymmetric systems. We demonstrate clock-pulling-enabled wireless power transfer operating at the maximum-efficiency state, establishing a paradigm that unifies non-Hermitian physics with modern control theory for practical WPT applications. The physics discussed in our work is generally applicable to non-Hermitian systems and may bring insights into engineering various platforms such as waveguide resonators \cite{iorsh2020waveguide,veenstra2024non}, acoustics cavities \cite{shao2020non,hu2021non}, optoelectronics \cite{zhang2018parity,chembo2019optoelectronic}, etc.

\section{Acknowledgements}
The author is grateful to Prof. Yu Chai for helpful discussion. The work is supported by the National Natural Science Foundation of China (52407015), the Postdoctoral Fellowship Program of CPSF (GZB20240469), the Sichuan University Interdisciplinary Innovation Fund, and the Shaanxi Province Key R\&D Program (2024GX-YBXM-236).

\bibliography{main} 

\end{document}